\documentclass[prl,twocolumn,showpacs,amsmath,amssymb,superscriptaddress]{revtex4-2}
\usepackage[utf8]{inputenc}
\usepackage{setspace}

\usepackage{graphicx}%Inchttps://www.overleaf.com/project/6007065131fba544597215eflude figure files
\graphicspath{ {figures/} }
\usepackage{lipsum} % for dummy text
\usepackage{wrapfig}% Have to use a package to do text wrapping
\usepackage{ifthen}% To use conditional command ifthenelse
\usepackage{color} % So that I can color the comment marks
\usepackage{bm} % bold math
\usepackage{multirow}% For tables
\usepackage{hyperref} % adds links
\usepackage{bbold} % allow for unit matrix symbol
\def\DFT{\textrm{DFT}}
\def\QP{\textrm{QP}}
\def\KS{\textrm{KS}}

\begin{document}

\title{Spin-flip Bethe-Salpeter equation approach for ground and excited states of open-shell molecules and defects in solids}
%\title{The Spin-Flip Bethe Salpeter Equation approach for transition energies of open-shell molecules and defects in solids}

\author{Bradford A. Barker}
\email{bbarker6@ucmerced.edu} 
\affiliation{Department of Physics, University of California,
  Merced, CA 95343, USA}
\author{David A. Strubbe}
\email{dstrubbe@ucmerced.edu}
\affiliation{Department of Physics, University of California,
  Merced, CA 95343, USA}

\date{\today}


\begin{abstract}
Open-shell systems such as magnetic molecules or defects with a triplet ground state are challenging to describe in electronic structure methods, but are of great interest for quantum information applications. We demonstrate a spin-flip approach within the Bethe-Salpeter equation to calculate ground and excited states of open-shell molecules and defected solids. The approach works in periodic boundary conditions without any need for embedding or selection of a subspace. Our benchmark results for the torsion potential-energy surface of ethylene and the optical excitations of the diamond NV$^{-}$ center show excellent agreement with the literature, and a low or moderate level of spin contamination.
\end{abstract}

%The spin-flip method, well-established in the theoretical chemistry literature, is commonly used to access open-shell states in conjunction with Time Dependent Functional Theory (SF-TDDFT). While SF-TDDFT inherits limitations from the DFT Kernel, SF-BSE calculates electron-hole interactions from the screened Coulomb interaction $W$, which can accurately capture charge-transfer and Rydberg states.

\maketitle

%%========================================================================
\textit{Introduction}.--Defects in solids have been a focus of great recent interest for quantum information. A canonical example is the diamond NV$^-$ center, consisting of substitutional N atom adjacent to a vacancy, and having an extra negative charge. This system has wide-ranging applications, from qubits\cite{PhysRevLett.93.130501,epstein2005anisotropic,childress2006coherent,dutt2007quantum,hanson2008coherent,fuchs2009gigahertz} to nanosensing\cite{doherty2014electronic,doherty2014temperature,dolde2014nanoscale,beams2013nanoscale}, and its optical properties are well studied both experimentally\cite{goss1996twelve,he1993paramagnetic,lenef1996electronic,davies1976optical,rogers2008infrared} and theoretically\cite{doherty2012theory,breuer1995ab,gali2019ab,choi2012mechanism,lischner2012first}. 
catalysts, and single-photon emitters
There has been a search for other defects with appropriate electronic structure, in bulk as well as two-dimensional materials recently, especially on the computational side \cite{Ivady, dreyer2018first, ping2021computational}. However, computing their electronic structure poses challenges due to their open-shell electronic configurations, meaning electrons singly occupy one or more energy levels.% in a molecule or defected solid. Excited-state methods such as $GW$ have been extended to compute the electron addition/removal excitation energies of open-shell systems \cite{lischner2012first}, but a $GW$/Bethe-Salpeter Equation (BSE) method to calculate neutral excitation energies for any configuration other than a singlet proves elusive.
The complications of such systems are well appreciated in the quantum-chemistry community \cite{krylov2006spin, casanova2020spin} but only recently have gathered attention in condensed-matter physics.

Density-functional theory (DFT) is the standard workhorse method for electronic structure, but it can only control for the total magnetic quantum number $M_S$ and not the total spin quantum number $S$, meaning that solutions are not necessarily eigenstates of $\hat{S}^2$ and therefore do not belong to a proper spin manifold (spin contamination). Further, DFT is single-reference, with the non-interacting Kohn-Sham (KS) system implying a single Slater determinant for the many-body wavefunction. Many investigations of open-shell defects rely on the simple but crude approximation of constrained DFT \cite{choi2012mechanism, Ivady,ciccarino2020strong,jin2021photoluminescence}, in which different ground or excited states are described by simply altering the occupations of single-particle KS states. There is not even a distinction between a singlet and triplet of $M_S=0$, a significant limitation in the ability to describe spin physics.

Multi-reference methods from quantum chemistry that explicitly account for the multiconfigurational character of the many-body wavefunction are applicable for molecular systems but are quite challenging to apply directly to solid-state systems \cite{wang2020excitons} including defects. Instead, many works have instead used embedding schemes \cite{sun2016quantum}, in which a finite subspace of in-gap defect states is selected, and then treated with configuration interaction or other correlated methods, in the presence of the solid environment \cite{ma2020quantum, bockstedte2018ab, pfaffle2021screened, bhandari2021multiconfigurational,choi2012mechanism}. Difficult questions of how to correct double-counting of correlation arise in such theories \cite{vorwerk2021quantum,muechler2022quantum}.

The spin-flip method \cite{krylov2001size} allows for the calculation of ground- and excited-state energies for systems with multiconfigurational character, such as open-shell systems or molecules breaking bonds, from a single-reference wavefunction.
The method computes excitation energies from a high-spin reference state, where $S = M_S$, to a set of lower-spin target states with $M_S - 1$, as diagrammed in Fig. S1 \cite{SM}.
%and an expectation value $\langle S^2\rangle$ which must be computed.
While DFT's Hohenberg-Kohn Theorem\cite{hohenberg1964inhomogeneous}
%, at the core of DFT, 
requires a non-degenerate ground state, DFT is able to compute the lowest-energy state for a given symmetry\cite{lischner2012first,cederbaum1974green}
%There are some number of partially-occupied open-shell states, some further number of closed-shell states, and some unoccupied virtual states.
%Next, we then generate a set of target states by removing some particular up-spin orbital, and then occupying a down-spin orbital.
The choice $S=M_S>0$ for the high-spin reference state is made throughout this work. %though further developments of the spin-flip approach include corrections from incorporating a second high-spin reference state with $S=-M_S$ \cite{lee2018eliminating}.
The target states with one spin flipped provide a basis of transitions.
%for the eigenstates of the Hamiltonian used, which captures the interaction between the electron-hole pair in these configurations.
Some target states may include effective double excitations with respect to the ground state, allowing for the description of conical intersections as per Brillouin's Theorem \cite{levine}.
%and bond dissociation.
%The spin-flip method can only capture some of the low-lying double-excitations, but this allows for calculation of difficult chemical scenarios such as bond dissociation.

The spin-flip method applied to TDDFT (SF-TDDFT) has demonstrated success for the description of antiferromagnetic molecules \cite{mayhall2014computational,mayhall2015computational}, conical intersections \cite{axelrod2022excited}, and ordering of near-degenerate ground- and excited-states. However, SF-TDDFT relies on the kernel $f_{xc}$ to describe the interactions between electrons and holes. Conventional options give incorrect long-range behavior \cite{onida2002electronic,wing2019comparing}. Moreover, there is a zoo of possible options and unclear which to choose \cite{shao2003spin, monino2021spin}. Collinear semilocal kernels, as would be typically used in TDDFT, do not mix different transitions, and so to be useful SF-TDDFT requires hybrid functionals with an unusually large fraction of exact exchange \cite{shao2003spin} or noncollinear functionals \cite{bernard2012general}.

Instead, in this Letter we draw on the many-body perturbation theory approach of $GW$-BSE \cite{hedin,Strinati,hybertsen1986electron, rohlfing2000electron, onida2002electronic} to develop the spin-flip Bethe Salpeter Equation approach (SF-BSE), 
%. The interaction between electrons and holes is described by the BSE kernel
%, which for spin-flipping transitions is just the direct term
with a kernel based on the screened Coulomb interaction $W$. This allows us to describe excitonic effects \cite{rohlfing2000electron}, valence excitations, Rydberg excitations, and charge-transfer excitations 
%from an improved description of the long-range behavior of the electron-hole interaction
\cite{blase2020bethe}. Our method works naturally in a periodic cell, like conventional $GW$/BSE, without the need for any selection of a subspace, embedding technique, or double-counting corrections.

Independent from the present effort \cite{barker2019spin}, Monino and Loos \cite{monino2021spin} recently considered an SF-BSE approach, applied to atoms and molecules with a detailed comparison to different quantum chemistry and TDDFT approaches. Our work marks a major advance in applying SF-BSE to defects in solids, demonstrating high accuracy for potential-energy surfaces, and identifying the problematic nature of open-shell $GW$ for use in BSE calculations.

%Combining the Spin-Flip method with the BSE, we have the SF-BSE Hamiltonian.
%\cite{monino2021spin}
%The Hamiltonian, expanded in a basis set of electron-hole pairs, has two terms:
%a simple single-particle energy eigenvalue difference between the electron, $E_c$, and the hole $E_v$, along the diagonal; and
%the interaction kernel.
%In the previously established Spin-Flip Linear-Response Time-Dependent Density Functional Theory, the interaction kernel
%is the exchange-correlation functional of TDDFT.
%However, in Spin-Flip Bethe-Salpeter Equation, the interaction kernel is the screened Coulomb interaction, $W$.
%The screened Coulomb interaction has the appropriate short- and long-ranged behavior for the interaction between the electron and hole.
%The Bethe-Salpeter Equation is the eigenvalue equation appropriate for an approach based on Many-Body Perturbation Theory,
%and it allows us to describe molecules as well as defected solids.

%%========================================================================

%%========================================================================
\textit{Methods}.--The single-particle orbitals and eigenvalues for the high-spin reference state $|\textrm{H.S.},\,\textrm{Ref}\rangle$ are calculated in spin-polarized (unrestricted) DFT. We use the PBE \cite{perdew1996generalized} exchange-correlation functional for real-space pseudopotential \cite{hamann2013optimized, van2018pseudodojo} calculations in Octopus \cite{octopus2015,octopus2020}, with parameters defined in \cite{SM}. For comparison, some spin-unpolarized (restricted) calculations are also done \cite{SM}.
%As shown in Fig. \ref{fig:SF_schematic}: suppose we have a spin-polarized system with $N=N_{\alpha}+N_{\beta}$ electrons such that $N_{\alpha} > N_{\beta}$. The high-spin reference state from DFT gives us a set of $\alpha$-spin KS orbitals $\phi_m$ and $\beta$-spin orbitals $\psi_n$, with $m \in \{ 1,\ldots,N_{\alpha}\}$ and $n \in \{ 1,\ldots,N_{\beta},N_{\beta}+1,\ldots,N_{\beta}+N_{\textrm{unocc}}\}$. The orbitals $n \in \{ 1,\ldots,N_{\beta}\}$ are always occupied; for the state $|N,I\rangle$, one of the orbitals $i \in \{ 1,\ldots,N_{\alpha}\}$  becomes unoccupied (counting down: $N_{\alpha}+1 - i$) while one of the orbitals $j \in \{N_{\beta}+1,\ldots,N_{\beta}+N_{\textrm{unocc}}\}$ becomes occupied (counting up: $N_{\beta}+j$).

%\begin{widetext}
%\begin{figure*}
%\centering
%\includegraphics[scale=0.5]{Spin_Flip_Schematic.pdf}
%\caption{Schematic of the spin-flip method. For SF-BSE, the spin-flip excitation is calculated via the Bethe-Salpeter equation.}
%\label{fig:SF_schematic}
%\end{figure*}
%\end{widetext}

The target states are the collection of the spin-flip excited states which preserve particle number, $|N,I\rangle$. As single excitations, these can be described with the Bethe-Salpeter equation, which can
%. If these excited states are single excitations, taking one occupied state and placing it in an unoccupied state, then the framework of the electron-hole excitations in the Bethe-Salpeter Equation from many-body perturbation theory applies.
%
%The wavefunctions for the $\textit{I}$-th excitation, $|N,I \rangle$, may be written in real-space, within the Tamm-Dancoff Approximation, as
%\begin{equation} \label{eq:exciton_wfns}
%\Psi^{\textit{I}}_{M_s -1}(\vr_e,\vr_h) = \sum_{vc}
% A^{\textit{I}}_{v\alpha,c\beta}\, \psi_{c\beta}(\vr_e)\psi^*_{v\alpha}(\vr_h).
%\end{equation}
%
be written in the usual Tamm-Dancoff approximation \cite{rohlfing2000electron} for the case of a spin-flip from $\alpha$ to $\beta$:
\begin{equation}
\left(E^{\QP}_{c\beta} - E^{\QP}_{v\alpha} \right)A^{I}_{v\alpha,c\beta}\, + \sum_{v',c'} K_{v'\alpha c'\beta,v\alpha c\beta} A^{I}_{v\alpha,c\beta} = \Omega^I A^{I}_{v\alpha,c\beta}.
\end{equation}
with quasiparticle energies $E^{\QP}$ for conduction states $c$ and valence $v$, an excitation energy $\Omega$, and an eigenvector $A$. The BSE kernel $K$ has direct and exchange terms, $K^D$ and $K^X$, respectively. %The matrix elements are
%\begin{widetext}
%\begin{align*}
%& K^D_{vc,v'c'} = -\int \d \vr_1 \,\d\vr_2\, \psi^*_{c}(\vr_1,\beta)\psi_{c'}(\vr_1,\beta)W(\vr_1,\vr_2;\omega=0) \psi_{v}(\vr_2,\alpha) \psi^*_{v'}(\vr_2,\alpha),\\
%& K^X_{vc,v'c'} = \int \d\vr_1\, \d\vr_2\, \psi^*_{c}(
%vr_1,\beta)\psi_{v}(\vr_1,\alpha)v(\vr_1,\vr_2)\psi_{c'}(\vr_2,\beta) \psi^*_{v'}(\vr_2,\alpha).
%\end{align*}
%\end{widetext}
As with the triplet excitations in usual BSE \cite{rohlfing2000electron}, the $K^X$ matrix elements must be zero for spin-flips due to the orthogonality of the $\alpha$ and $\beta$ spinors.

Single-particle excitations of the high-spin reference state, in general, have multiple quasiparticle energies $E^{\QP}$ for a given electron addition/removal \cite{lischner2012first}. This issue has been neglected in some previous calculations \cite{ma2010excited}. It remains an open question how such quasiparticle multiplets (no longer indexable by $\alpha, \beta$) should be correctly used in BSE calculations, an issue which is not addressed in the SF-BSE work of Monino and Loos \cite{monino2021spin}. Errors on the order of the multiplet splittings ($\sim1-2$ eV in molecules and defects \cite{lischner2012first}) could result.
Using quasiparticle energies from the Generalized Plasmon Pole (GPP) approximation\cite{hybertsen1986electron} sidesteps this difficulty %by artificially enforcing that each electron addition/removal has only one quasiparticle peak,
but at the cost of ignoring the correct effects of the multiplet splittings \cite{lischner2012first}. Instead we take the pragmatic approach of simply using the KS energy eigenvalues $E^{\KS}$ instead of $E^{\rm QP}$. (By contrast, neglecting the kernel produces completely wrong results \cite{SM}.)
%\begin{equation}
%\left(\epsilon^{\KS}_{c\beta} - \epsilon^{\KS}_{v\alpha} \right)A^I_{v\alpha,c\beta}\, + \sum_{v',c'} K_{vc,v'c'} A^{I}_{v'\alpha,c'\beta} = \Omega^I A^I_{v\alpha,c\beta}.
%\end{equation}
The ground-state energy is the sum $E^{\DFT}_{\rm tot}(\textrm{Ref}) + \Omega^{I_{\rm min}}$, using the spin-flip state with the lowest energy, possibly negative. As we will see, the gap correction due to $GW$ is mostly just a constant offset \cite{SM}, which does not affect differences between various values of $\Omega^{I}$.

%The Kohn-Sham energies for SF-BSE@DFT may be those computed directly from DFT, or they may be shifted according to their updated values from further first-principles calculations, for instance a quasiparticle energy calculation with a plasmon-pole model \cite{hybertsen1986electron}, which will exclude the spin information of the one-particle excited state, or some suitable average of spin-split quasiparticle energies found from an open-shell $GW$ calculation \cite{lischner2012first}. To the bare Kohn-Sham energy eigenvalue $\epsilon_{n}^{\DFT}$ is added some shift $\Delta_{n}$, which will adjust the values of the diagonals for the SF-BSE@DFT Hamiltonian.

%[Talk about where parameters come from in the screened Coulomb interaction.]

%\begin{widetext}
\begin{figure}
\centering
\includegraphics[scale=0.5]{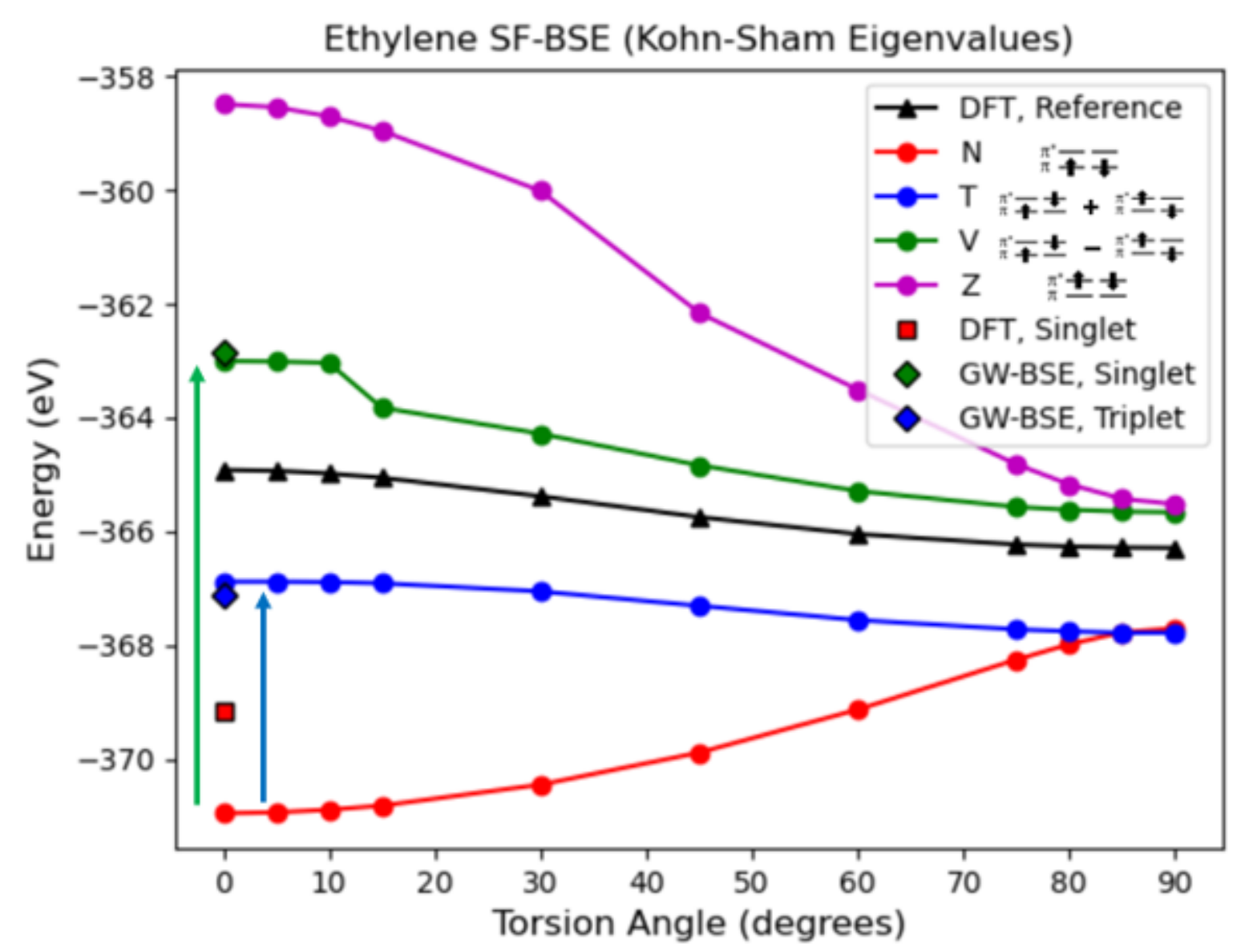}
\caption{The potential-energy surfaces of the N, T, V, and Z states for ethylene under torsion, listed with their main Slater determinants at $0^{\circ}$. The high-spin DFT reference energy is shown in black.
The DFT ground-state (singlet) energy and lowest triplet and singlet excitations from conventional $GW$-BSE are shown at $0^{\circ}$.}
\label{fig:NTVZ}
\end{figure}
%\end{widetext}

%%========================================================================

\textit{Ethylene}.--Ethylene is a well-studied system, both for vertical transition energies from its ground-state and for its torsion potential, and is a classic test for spin-flip methods \cite{shao2003spin}. The torsion potential %estimates the amount of energy required to break the ethylene $\pi$ bond by 
twists the $\pi$ bond,
%one H-C-H unit about the $\sigma$-bond axis to $90^{\circ}$. This torsion modifies the electronic structure of the molecular orbitals so that the 
making the $\pi$ and $\pi^*$ orbitals become degenerate at $90^{\circ}$\cite{SM}, making it an open-shell system with a triplet ground state.

% Put me in SM
%We use Optimized Norm-Conserving Vanderbilt pseudopotentials \cite{hamann2013optimized} from the Pseudo-Dojo pseudopotential database \cite{van2018pseudodojo}, The relaxed atomic coordinates (at $0^{\circ}$) were calculated with a 0.115~{\AA} real-space grid spacing, equivalent to roughly 115 Ry planewave wavefunction cutoff, in a box with edge-length 12~{\AA}. A smaller box size was used in the subsequent calculations, which contains 99\% of the charge density for both the ethylene molecule with no torsion and $90^{\circ}$ of torsion.
%The ground- and excited-state energies were calculated explicitly at torsions of 0, 5, 10, 15, 30, 45, 60, 75, 80, 85, and 90$^{\circ}$. For DFT input to SF-BSE calculations, we used a real-space grid spacing for both the wavefunctions and density of 0.34 Bohr, equivalent to about 85 Ry planewave cutoff for the wavefunctions. We obtained the $M_S=1$ reference state by constraining occupations. Based on \cite{van2015gw}, we used 860 empty states and 24~{Ry} for the calculation of the screened Coulomb interaction.
%[Optional information about frequency dependence: Smearing of 0.2 eV, frequency steps of 0.2 eV. Low frequency cut of 115 eV. Cite Mauro and others if needed.]
We obtained the $M_S=1$ reference state by constraining occupations. We performed extensive tests of convergence, finding similar characteristics to ordinary BSE, and demonstrating how SF-BSE does not require \textit{a priori} knowledge to select an appropriate subspace.

The N, T, V, and Z potential energies for ethylene under torsion, listed with their highest contributing Slater determinants at zero torsion angle, are shown in Fig. \ref{fig:NTVZ}. %The potential barriers are excitations from the high-spin reference state DFT energy, shown in black. Also at zero torsion are shown the DFT ground-state (singlet) energy, and lowest triplet and singlet optical excitations as computed from conventional $GW$-BSE.
%The N $\rightarrow$ T transition energy computed from SF-BSE is 4.07~{eV}, from conventional $GW$-BSE, 3.85~{eV}, with experimental values ranging from 4.21 $-$ 4.68~{eV} \cite{van1976low}. The N $\rightarrow$ V transition energy computed from SF-BSE is 7.95~{eV}, from conventional $GW$-BSE, 8.09~{eV}, with an experimental value estimated to be 7.88~{eV} \cite{wu2015v}.
The excitation energies computed within $GW$-BSE required 500 empty states for the BSE kernel to achieve convergence, while SF-BSE converges with about 55 empty states \cite{SM}. We note that the SF-BSE Hamiltonian was not constructed using quasiparticle energies, while displaying remarkably high agreement with the (conventional) $GW$/BSE calculations (Table \ref{tab:ethylene_transitions}). This is an important demonstration of consistency with the usual $GW$/BSE. Examination of the relation between KS eigenvalues of the high-spin reference and the closed-shell ground state shows that they actually differ by the quasiparticle corrections \cite{SM}, emulating $GW$. This is suggestive of studies such as \cite{kraisler2014fundamental} in which a local functional, used in a particular way, is able to provide a derivative discontinuity to correct the gap. We note that by contrast neglecting quasiparticle energies with conventional BSE gives completely wrong excitation values of -3.00 and 1.27~{eV} for N $\rightarrow$ T and N $\rightarrow$ V, respectively. Our results also are in good agreement with experiment and with quantum Monte Carlo. Using GPP at $0^{\circ}$ also provides similar results \cite{SM}. 
%Further studies should reveal whether SF-BSE can indeed generally provide accurate and more easily converged optical excitations of molecules.

%We use the T excitation as our estimate of the triplet energy to treat states on an equal footing \cite{krylov2006spin,shao2003spin}; it has a constant energy offset vs. DFT. 
Our N state can be compared with the usual DFT ground state at $0^{\circ}$, and is slightly lower which seems to be due to improving static correlation from mixing of a doubly-excited configuration. This energy difference
%(much larger in restricted calculations, Fig. S3 \cite{SM})
is comparable conceptually and numerically to the differences in estimating atom energies by $GW$ as HOMO of a neutral atom or LUMO of a cation \cite{bruneval2012ionization}. We find that restricted calculations for ethylene are generally worse \cite{SM}, as for molecules in \cite{monino2021spin}.

\begin{table}[h!tbp]
\begin{center}
\begin{tabular}{c|c|c|c|c } 
\hline
Transition & $GW$-BSE & SF-BSE & SF-$GW$-BSE & 
%VMC & DMC\cite{wu2015v} & 
Experiment \\
\hline \hline
N $\rightarrow$ V & 8.09 & 7.95 & 8.01 &
%7.96 & 7.93 &
7.88\cite{wilkinson1955far,davidson1998zero} \\
N $\rightarrow$ T & 3.85 & 4.07 & 4.19 & 
%N/A & N/A &
4.21--4.68\cite{van1976low} \\
\hline
\end{tabular}
\caption{Transition energies, in eV, for the singlet N$\rightarrow$ V and triplet N$\rightarrow$ T transitions, from conventional $GW$-BSE, SF-BSE, and experiment. Additional results for N $\rightarrow$ V are 7.96 eV from variational Monte Carlo and 7.93 eV from diffusion Monte Carlo \cite{wu2015v}.}
\label{tab:ethylene_transitions}
\end{center}
\end{table}

\begin{figure}
\centering
\includegraphics[scale=0.4]{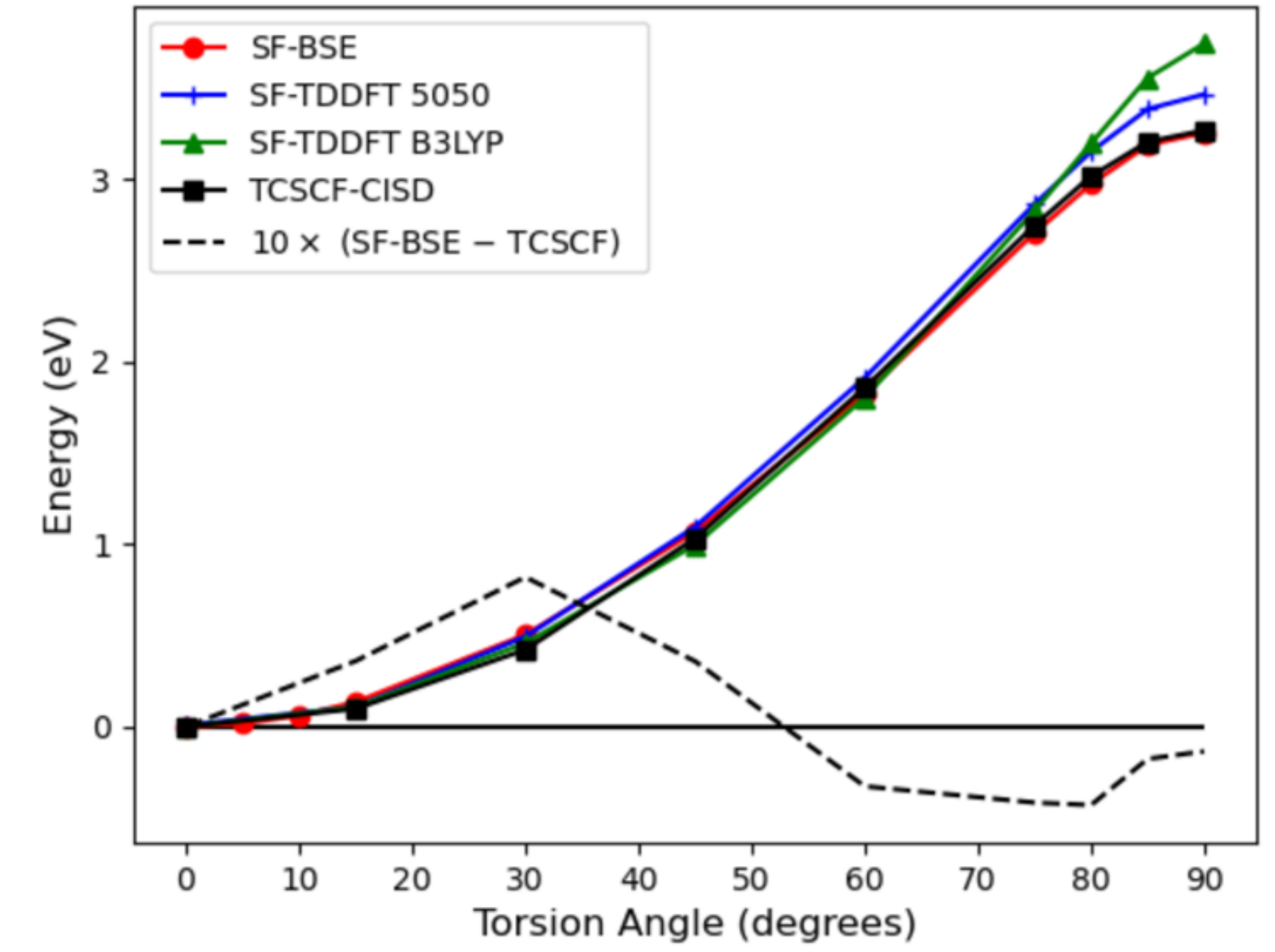}
\caption{The ethylene torsion potential, comparing SF-BSE to SF-TDDFT and TCSCF-CISD results from \cite{shao2003spin}. The dashed curve is the difference between the SF-BSE and the TCSCF-CISD results, multiplied by ten to be seen more clearly.}
\label{fig:ethylene_torsion_gs}
\end{figure}

The torsion potential is in excellent agreement with the ``gold standard'' Two-Configuration Self-Consistent Field-CISD \cite{shao2003spin}, and far closer than SF-TDDFT methods (Fig. \ref{fig:ethylene_torsion_gs}). The barrier, which is the energy difference between the singlet N state at $90^{\circ}$ torsion and no torsion, is computed in SF-BSE as 3.25~{eV} and is
%The SF-TDDFT torsion barriers are 3.74 and 3.46~{eV}, with the B3LYP and 50-50 hybrid functionals, respectively \cite{shao2003spin}.
%The torsion barrier calculated from the ``gold standard'' Two-Configuration Self-Consistent Field-CISD (``TCSCF-CISD'') is
3.27~{eV} in TCSCF-CISD \cite{shao2003spin}.
%The TCSCF-CISD is expected to be similar to that of a Complete Active Space SCF method.
%Further, we show the evolution of the $N$-state potenti from 0 to 90 degrees; at 0 degrees, the $G$ determinant has [value... 0.98 or so?] and $D$ has [value... 0.2?], while at 90 degrees, $G$ and $D$ have nearly equal weight.
We note that using GPP for the potential-energy surfaces produces significantly worse results \cite{SM}.

%\begin{widetext}
%\end{widetext}

\begin{figure}
\centering
\includegraphics[scale=0.4]{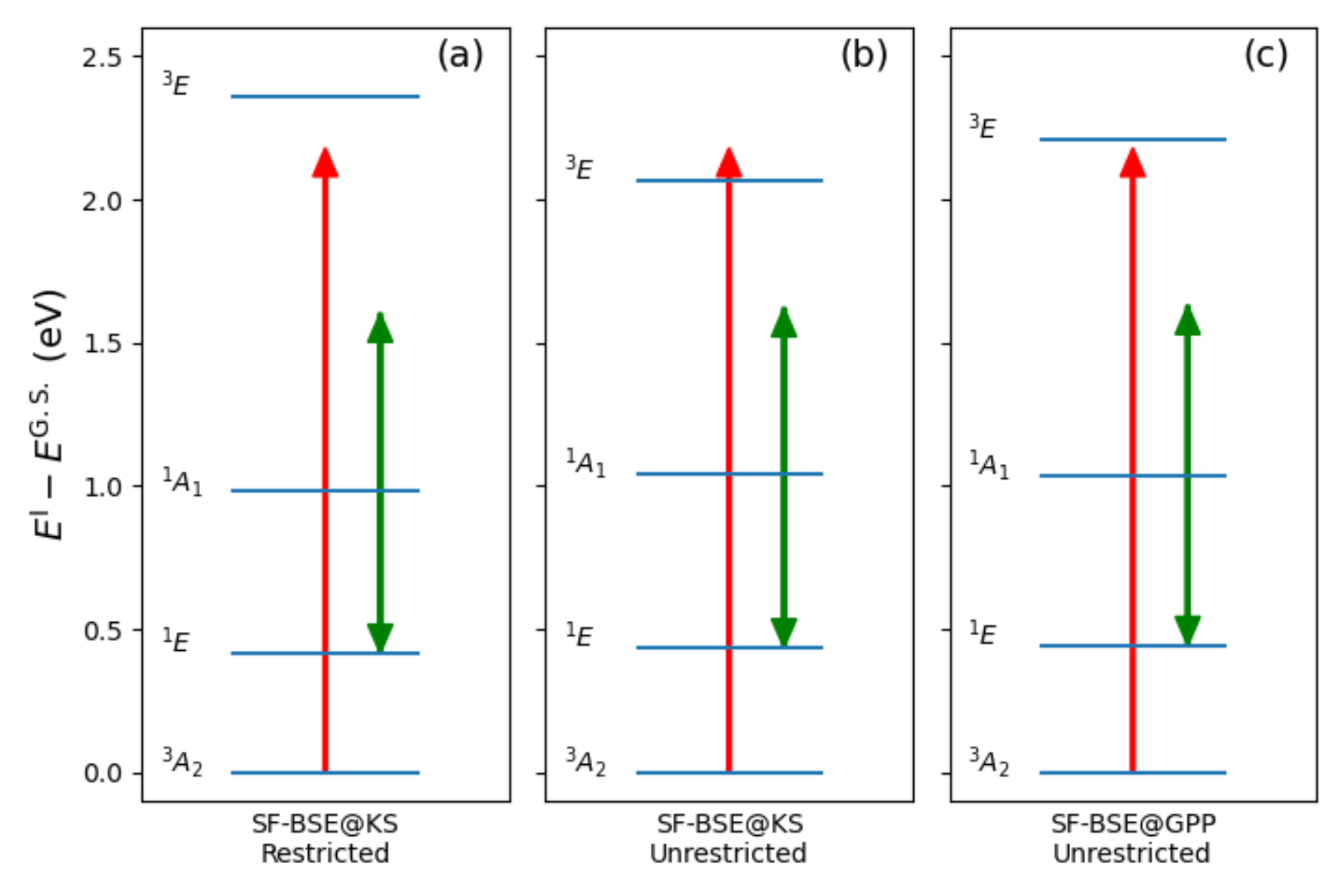}
\caption{Energies of the in-gap many-body defect states for the NV$^{-}$ defect in diamond, computed with SF-BSE, with (a) restricted KS eigenvalues, (b) unrestricted KS eigenvalues, and (c) unrestricted GPP quasiparticle energies  \cite{hybertsen1986electron}. The red arrow is the experimental value  \cite{davies1976optical} for the vertical optical transition, and the green double-headed arrow is the experimental value for the singlet splitting  \cite{rogers2008infrared}.}
\label{fig:NV_all_sfbse}
\end{figure}

\textit{NV\texorpdfstring{$^{-}$}{}  center in diamond}.--To demonstrate the applicability to defects, we consider the NV$^{-}$ center.
%This system has wide-ranging applications, from qubits\cite{PhysRevLett.93.130501,epstein2005anisotropic,childress2006coherent,dutt2007quantum,hanson2008coherent,fuchs2009gigahertz} to nanosensing\cite{doherty2014electronic,doherty2014temperature,dolde2014nanoscale,beams2013nanoscale}, and its optical properties are well studied both experimentally\cite{goss1996twelve,he1993paramagnetic,lenef1996electronic,davies1976optical,rogers2008infrared} and theoretically\cite{doherty2012theory,breuer1995ab,gali2019ab,choi2012mechanism,lischner2012first}.
%
% put me in SM
There is no need for constraining occupations, as the DFT and true ground states are triplets. 
%gives $M_S=1$ and the true ground state is a triplet, and the spin flip can be considered a physical process rather than a trick.
We calculate zero-phonon line (ZPL) transition energies from the total-energy differences $E^I - E^{\textrm{G.S.}}$ in SF-BSE. A comparison between the in-gap defect states computed with restricted and unrestricted KS orbitals is in Fig. \ref{fig:NV_all_sfbse}. Results using KS eigenvalues have an appreciable difference only for the excitation energy to $^3E$, 0.30~{eV}. Unrestricted KS and GPP results  (with 300 empty states in the Coulomb-hole sum \cite{BerkeleyGW}) are fairly close.

\begin{figure*}
\centering
\includegraphics[scale=0.5]{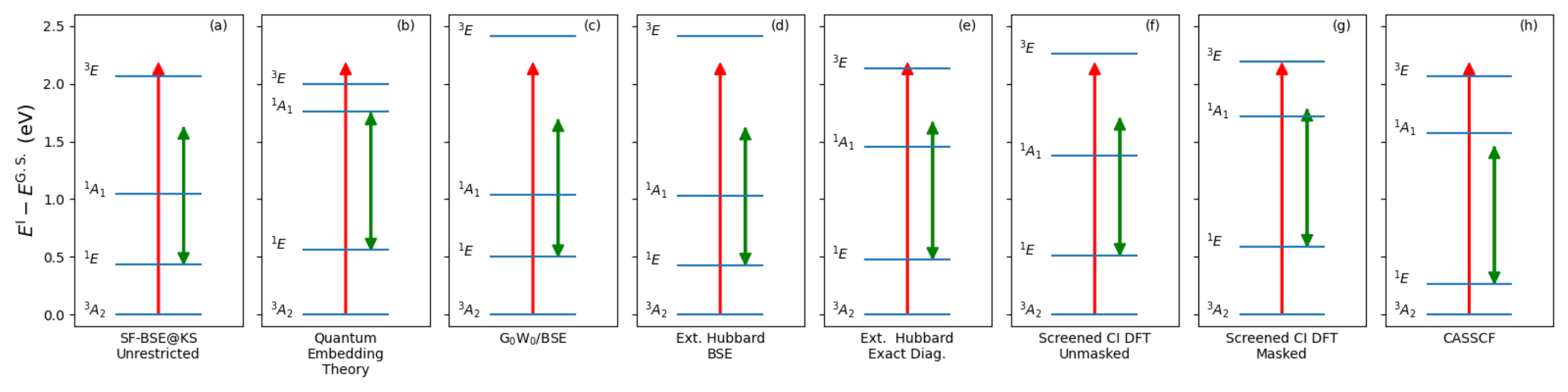}
\caption{Energies of the in-gap many-body defect states for the NV$^{-}$ defect in diamond, computed with SF-BSE compared to various other results in the literature: (b) \cite{ma2020quantum}, (c) \cite{ma2010excited}, (d) and (e) \cite{choi2012mechanism}, (f) and (g) \cite{pfaffle2021screened}, and (h) \cite{bhandari2021multiconfigurational}. The red arrow is the experimental value\cite{davies1976optical} for the vertical optical transition, and the green double-headed arrow is the experimental value for the singlet splitting\cite{rogers2008infrared}.
}
\label{fig:NV_literature}
\end{figure*}

A comparison between 
%the excitation energies computed from
SF-BSE and other methods is presented in Fig. \ref{fig:NV_literature}.
%demonstrating excellent agreement with experiment for excitation to $^3E$.
The $^1A_1$ excitation energy shows appreciable difference from experiment in SF-BSE and other methods that cannot include a contribution from the determinant in which the up- and down-spin $v$ orbitals are empty and the $e_x$ and $e_y$ orbitals are occupied, unlike
%. Methods that do allow for this multiply-excited determinant show much improved agreement with experiment 
\cite{ma2020quantum,bockstedte2018ab,pfaffle2021screened,bhandari2021multiconfigurational}. This state would require a double spin-flip excitation from our $M_S = 1$ reference state.
%, but could be accessed from an alternate reference state with different occupations.

%%========================================================================
\textit{Spin contamination}.--
Excited-state theories, especially spin-flip methods, encounter the problem of spin contamination, which can be a diagnostic of poor energies or wavefunctions, or also simply cause difficulty in identifying the spin character of the calculated states. The eigenvector describing the excited state yields an expectation value of $\hat{S}^2$ that may differ from proper values of $S\left(S+1\right)$ such 0 (singlet) or 2 (triplet). Sources of spin contamination include the difference in spatial wavefunctions for the different spin channels for spin-polarized DFT \cite{baker1993spin}, or the absence of particular transitions from the set of target states that are required to form an eigenvector of $\hat{S}^2$ \cite{lee2018eliminating}. Projection methods can be used to correct spin contamination and energies \cite{yamaguchi1988spin,tada2021estimation}.
%, though they are not used in the present work.

%After having evaluated SF-BSE eigenvectors $|\Psi^I\rangle$, analysis of their nature must commence. Beyond identifying the contributions of particular single-particle spin-flip transitions to the eigenvector, the value of $\langle \hat{S}^2 \rangle $ must be identified, as the excitation is only guaranteed to have $M_S$ fixed to $M^{\textrm{H.S.}}_S -1 $, with $M^{\textrm{H.S.}}_S$ the $M_S$ quantum number of the initial High-Spin Reference state.
%When values of $\langle \hat{S}^2 \rangle $ for some excitation eigenvector $|\Psi^I\rangle$ do not obey exactly $\langle \hat{S}^2 \rangle = S(S+1) $ for positive integer $S$, there is some source of spin contamination.

%In a ``spin-restricted'' approach, the orbitals or bands have the same spatial wavefunction in both the up- and down-spin channels. In such an approach,
%Within a spin-restricted approach to constructing the orbitals for both up- and down-spin channels,
%the orbitals themselves do not introduce spin-contamination \cite{wang1995evaluation}. The eigenvectors, however, may be ``incomplete'', missing a contribution from a spin-down to spin-up transition, or a linear combination of transitions that belong to $|S,M_S - 1\rangle$ and $|S-1,M_S -1\rangle$ spaces.
%In this latter case, pure $S$ and $S-1$ states may be obtained by forming linear combinations from different excitation eigenvectors using the appropriate Clebsch-Gordan coefficients, for both the states and the excitation energies.

\begin{figure}
\centering
\includegraphics[scale=0.5]{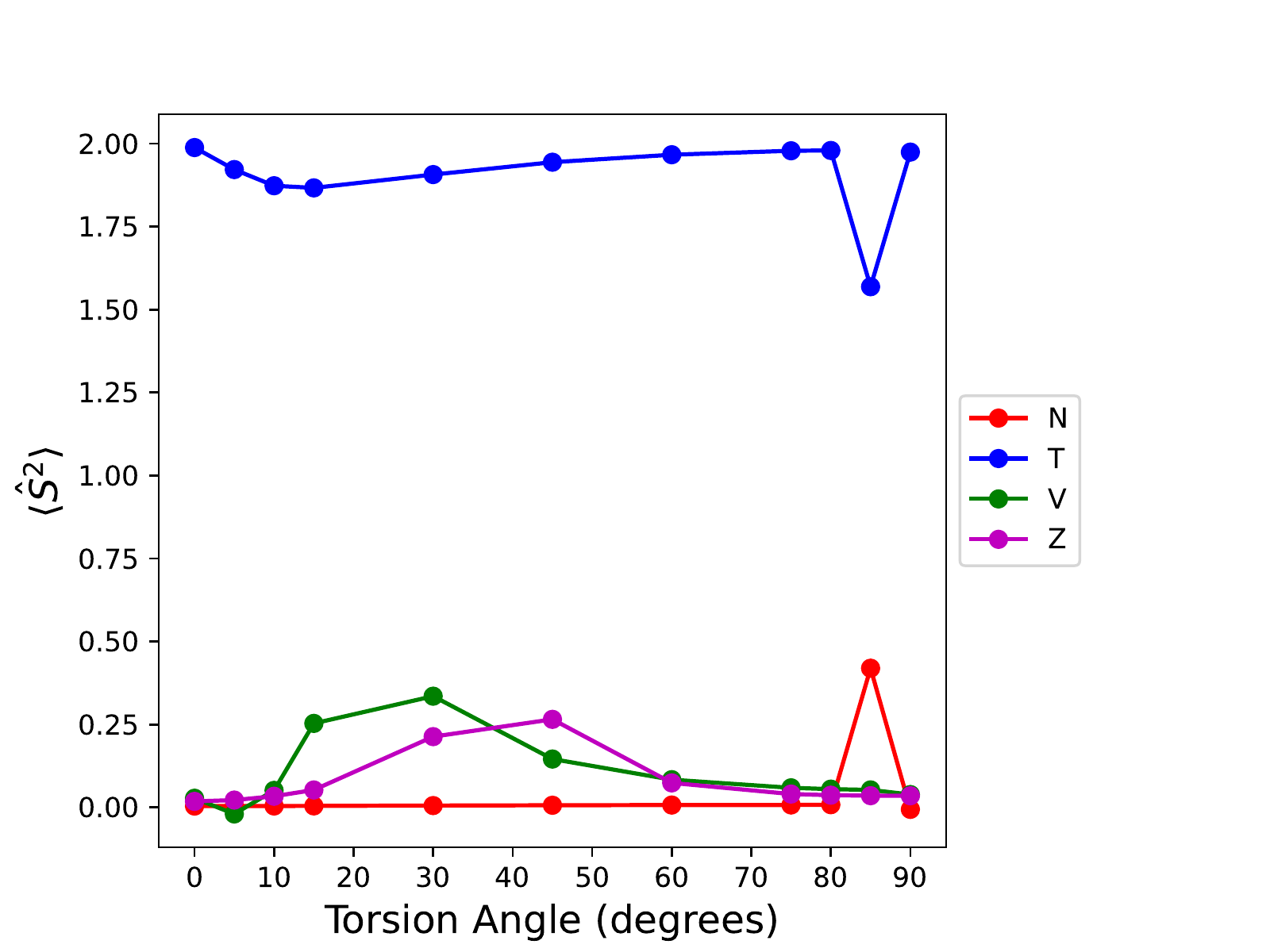}
\caption{The computed values of $\langle \hat{S}^2 \rangle$ for the N, T, V, and Z states of ethylene within SF-BSE. T is a triplet while the others are singlets.}
\label{fig:spin_contamination}
\end{figure}

%With spin-unrestricted Kohn-Sham, however, the orbitals allow for additional spin contamination, and 
If a many-body wavefunction is available, the the L{\"o}wdin formula \cite{lowdin1955quantum, SM} can be used to compute $\langle N, I | \hat{S}^2 | N, I \rangle$ for state $I$.
%In DFT, we do not have a many-body wavefunction, but 
L{\"o}wdin results using the KS Slater determinant have been found to be similar to those from density functional approximations to $\langle \hat{S}^2 \rangle$ \cite{wang1995evaluation}, and we find results very close to 2 for our high-spin reference states \cite{SM}. Within an excited-state theory, it is natural to evaluate $\langle N, I | \hat{S}^2 | N, I \rangle$ via
%be considered to analyze the spin quantum numbers of the eigenvectors. In Refs. \cite{ipatov2009excited} and \cite{myneni2017calculation}, the ``superoperator'' formalism is employed to compute instead
the difference $\Delta \langle \hat{S}^2 \rangle_I$ from the ground state value,
%\begin{align}
%\Delta \langle \hat{S}^2 \rangle_I = \langle N, I | \hat{S}^2 | N, I \rangle - \langle N, \textrm{H.S. Ref} | \hat{S}^2 | N, \textrm{H.S. Ref} \rangle
%\end{align}
which can be computed from only the excitation eigenvector.
%, as $\langle N, \textrm{H.S. Ref} | \hat{S}^2 | N, \textrm{H.S. Ref} \rangle$ is readily computed from the well-known L{\"o}wdin Formula\cite{lowdin1955quantum}.
%Within the superoperator formalism, the spin-flipping operator $\hat{\mathcal{O}}^{\dagger} = a^{\dagger}_{\beta}i_{\alpha}$ generates $\Delta \langle \hat{S}^2 \rangle_I$ from $\langle \Psi^{\textrm{H.S. Ref}} | \left[ \hat{\mathcal{O}} , \left[  \hat{S}^2, \hat{\mathcal{O}}^{\dagger} \right] \right] |  \Psi^{\textrm{H.S. Ref}} \rangle$ .
The result for $\Delta \langle \hat{S}^2 \rangle_I$ with a spin-flip (different from the spin-conserving expression) is given in \cite{li2011spin} in the context of SF-TDDFT, and detailed in our Supplementary Material \cite{SM}. The related spin-conserving expression of \cite{ipatov2009excited} could be used to for $GW$-BSE in the presence of spin-orbit coupling \cite{barker2022fully}. We note that in TDDFT such expressions are an approximation, based on treating the KS orbitals as a Slater determinant \cite{ipatov2009excited}. By contrast, in BSE, we can rigorously calculate $\Delta \langle \hat{S}^2 \rangle_I$ consistently with the BSE derivation, as noted in \cite{pokhilko2021evaluation}, since only the exciton wavefunctions are needed \cite{Strinati,rohlfing2000electron}.
%BSE is an equation for the electron-hole amplitudes \cite{Strinati}, also known as the exciton wavefunctions \cite{rohlfing2000electron}, and these are the only quantities required for the two-body operator $\Delta \hat{S}^2$. As such, assessment of spin contamination is only approximate in SF-TDDFT, but is rigorous in SF-BSE.

We show results of $\langle \hat{S}^2 \rangle_I$ in Fig. \ref{fig:spin_contamination}. We find a small spin contamination comparable to TDDFT  \cite{lee2018eliminating} that still allows identifying singlet and triplet states. Spin contamination is substantially larger in NV$^-$ \cite{SM}, as others have found for solids \cite{tada2021estimation}. %This effect of incompleteness of the accessible transitions from our reference state,
States where the needed transitions are available still have values of $\langle \hat{S}^2 \rangle_I$ that enable identification of singlet and triplets.

%\cite{damascelli2003angle}

%%========================================================================
\textit{Conclusion}.--We have presented the SF-BSE method to calculate electronic structure for open-shell molecules and defects in solids. The ethylene torsion potential shows high agreement with a high-level multi-reference method, considerably improved from SF-TDDFT. Due to the problematic nature of $GW$ multiplets for BSE, we use KS energies as input. The optical transition energies in the equilibrium geometry show remarkable agreement with conventional $GW$-BSE and experiment, while requiring roughly one-tenth of the empty orbitals and no single-particle quasiparticle energies. This surprisingly successful approach may be promising for optical calculations of closed-shell systems. The in-gap defect states for the NV$^{-}$ in diamond show good agreement with experiment and multi-reference methods, apart from the too-low $^1A_1$ state. %The NV$^{-}$ results also indicate little effect from the use of quasiparticle energies in the SF-BSE Hamiltonian. 
Spin contamination is small for ethylene and manageable for NV$^-$. This method will be available in a future release of BerkeleyGW.

Our results show SF-BSE is promising for potential-energy surfaces in defects \cite{choi2012mechanism}. This method, unlike embedding schemes, can easily provide forces through BSE excited-state forces schemes \cite{Sohrab2003, strubbe_thesis, rafael2022} for efficient exploration and investigation of Stokes shifts \cite{Alkauskas_2014}, Jahn-Teller distortions \cite{abtew2011dynamic,ciccarino2020strong}, and internal conversion \cite{choi2012mechanism}.
%
%It is still an open question as to the proper treatment of $GW$ within SF-BSE, as certain single-particle excitations have multiplet solutions.
%The formalism presented in the supplementary material for the calculation of $\langle \hat{S}^2 \rangle$ for the excitations in SF-BSE can also be extended in future work to assist with analysis of excitations in conventional $GW$-BSE in the presence of spin-orbit coupling \cite{barker2022fully}, as these excitations are no longer eigenvectors of $\hat{S}^2$.
%
Other future directions for the SF-BSE formalism includes use of quasiparticle multiplets \cite{lischner2012first} or spinors \cite{barker2022fully}, multiplet reference states \cite{lee2018eliminating, li2011spin} %to generate spin-complete excited states, and use of 
or other reference states to access different excited states.

%%========================================================================
The authors thank Hrant Hratchian and Abdulrahman Zamani for useful discussions %about spin-contamination.
and John Parkhill for originally suggesting the concept
%of spin-flip Bethe-Salpeter to D.A.S. 
This work was supported by the U.S. Department of Energy, Office of Science, Basic Energy Sciences, CTC and CPIMS Programs, under Award DE-SC0019053.
Computational resources were provided by the National Energy Research Scientific Computing Center (NERSC), a U.S. Department of Energy Office of Science User Facility operated under Contract No. DE-AC02-05CH11231, and by the Multi-Environment Computer for Exploration and Discovery (MERCED) cluster at UC Merced, funded by National Science Foundation Grant No. ACI-1429783.

%%========================================================================

%\bibliography{references_spin_flip_bse}


\end{document}